\documentclass{aa}
\newcommand{\mincir}{\raise
-2.truept\hbox{\rlap{\hbox{$\sim$}}\raise5.truept\hbox{$<$}\ }}
\newcommand{\magcir}{\raise
-2.truept\hbox{\rlap{\hbox{$\sim$}}\raise5.truept\hbox{$>$}\ }}
\newcommand{\minmag}{\raise
-2.truept\hbox{\rlap{\hbox{$<$}}\raise6.truept\hbox{$<$}\ }}
 
\usepackage{graphicx}

\begin{document}

\thesaurus{11.03.1;11.04.1;11.05.02;11.06.2;11.09.2} 

\title{The Richness-Shape Relation of the 2dFGRS Groups}
\author{M. Plionis
\inst{1,2}
\and
 S. Basilakos
\inst{1}
}
\institute{Institute of Astronomy \& Astrophysics, National Observatory of
Athens, Palaia Penteli 152 36, Athens, Greece,
\and
Instituto Nacional de Astrof\'{i}sica \'Optica y
Electr\'onica, AP 51 y 216, 72000, Puebla, Pue, M\'exico
}

\titlerunning{The Shape of the 2dF Groups of Galaxies}
\authorrunning{M. Plionis \& S. Basilakos}

\date{Received ...2003 / Accepted  .. 2003}

\maketitle

\abstract{ 
We estimate the group shape of the recently 
compiled Pecolation-Inferred Galaxy Group (2PIGG) 
catalogue. Using a set of integral equations
we invert the projected distribution of group axial ratios
and recover the corresponding intrinsic distribution 
under the assumption that groups are pure spheroids.
In agreement with the analysis of the  UZC-SSRS2 group sample 
(Plionis, Basilakos \& Tovmassian 2004) we find that groups are extremely
elongated prolate-like systems.
We also find an interesting trend
between shape and group richness with poorer groups being 
significantly flatter than richer ones.

}

\section{Introduction} 
Groups of galaxies are the lowest level cosmic structures, after
galaxies themselves, in the hierarchy  that leads to the
largest  virialized structures, the clusters
of galaxies. It appears that most galaxies are found in groups and
they are therefore extremely important in our attempts to understand
the cosmic structure formation processes.
Since virialization will tend to sphericalize initial anisotropic
distributions of matter, the shape of different cosmic structures is
a possible indication of their evolutionary stage.

Theoretical and observational works have shown that the cosmic
structures (clusters and superclusters) 
are dominated by prolate like shapes (cf. 
Carter \& Metcalfe 1980; Plionis, Barrow \& Frenk 1991; 
Cooray 1999; Basilakos, Plionis \& Maddox 2000;
Zeldovich, Einasto \& Shandarin 1982;
de Lapparent, Geller \& Huchra 1991;
Plionis, Jing \& Valdarnini 1992;
Jaaniste et al. 1998; Sathyaprakash et al. 1998; 
Valdarnini, Ghizzardi \& Bonometto 1999; Basilakos, Plionis \&
Rowan-Robinson 2001).
In the case of Groups of galaxies a recent study of the UGC-SSRS2
group catalogue (Plionis, Basilakos \& Tovmassian 2004)
have shown that they are extremely flat systems.
Furthermore, their intrinsic shape appears to be that of a
prolate-like spheroid (see also Oleak et al. 1995 and references therein). 

In this letter we use the recently constructed
2PIGG group catalogue (Eke et al. 2004) which
is based on the Two-Degree Field Galaxy Redshift Survey 
(2dFGRS) and contains $7020$ groups with four or
more galaxy members to estimate their projected and intrinsic shape as a
function of their galaxy membership.

\section{Sample and Shape Determination Method}
The 2PIGG group catalogue (Eke et al. 2004) 
is constructed using a friends-of-friends (FOF)
algorithm with linking parameters selected after thorough
tests that have been applied on mock $\Lambda$CDM galaxy catalogues.
The underline 2dFGRS catalogue used contain 191440 galaxies with well
defined completeness limits for the survey geometry, the magnitude and redshift
selection functions. The resulting 2PIGG group catalogue contains 7020
groups with at least 4 members having a median redshift of 0.11.
Out of these 2975 and 4045 are found in the northern (NGP) and the
southern (SGP) regions of the survey respectively.

The specific group finding algorithm used (Eke et al. 2004)
treats in detail many issues that are related to completeness,
the underlying galaxy selection function and the resulting biases 
that enter in attempts to construct unbiased group or cluster catalogues.
In order to take into account the drop of the underlying galaxy number
density with redshift, due to its magnitude limited nature, the
authors have used a FOF linking parameter that scales with
redshift. This scaling is variable also in the perpendicular and
parallel to the line-of-sight direction with their ratio being $\sim
11$ (for details see the original paper of Eke et al. 2004).
The necessity to increase the linking volume with
redshift introduces biases in the
morphological and dynamical characteristics of the resulting groups
which should be taken into
account before extracting any statistical information from the group
catalogue. For example, an outcome of the above is the increase with
redshift of both the velocity dispersion and the projected size
of the candidate groups. In Fig.1 we present for the NGP and SGP
samples with membership $n_m\ge 4$ their velocity dispersions
and the mean projected member galaxy separation. 
The redshift dependence is evident.
Therefore, the
probability that the groups found are real dynamical entities, should
decrease with redshift. 
The best probably approach to deal with such systematics
is the use of N-body simulations to test the effects of the algorithm
and the parameters used as well as to calibrate the statistical
results (Eke et al. 2004). 

For our study of the shapes of the 2PIGG groups we will also follow
some standard procedures to minimize possible systematics.
To this end we extract group subsamples 
the number density of which, within some limiting redshift, is 
relatively constant, ie., we select a roughly volume limited region. 
In Fig. 2 we show the group number density as a function of redshift in
equal volume shells, which is roughly
constant out to $cz\sim 30000$ km s$^{-1}$. Within this limit 
we are left with 2980 groups
with 4 or more members (1493 and 1487 in the NGP and SGP
respectively).
\begin{figure}
\includegraphics[width=8.4cm]{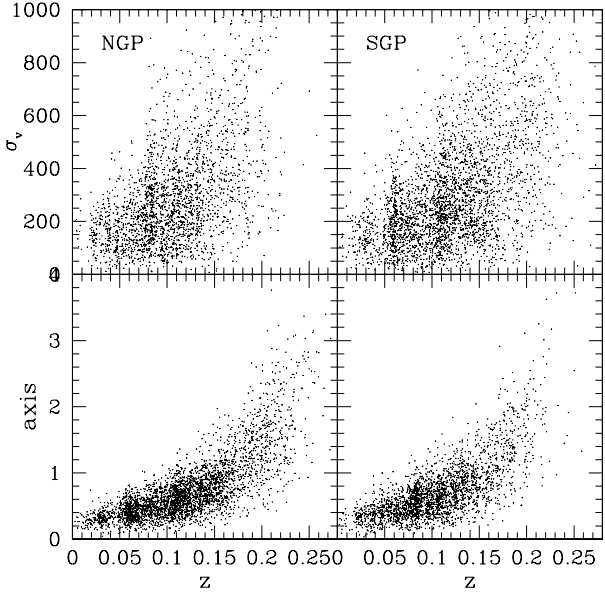}
\caption{The dependence of the group velocity dispersion (upper panels)
and the mean member intergalaxy separation
(lower panels) on redshift for the NGP and SGP 
samples.}
\end{figure}

\begin{figure}
\includegraphics[width=8.4cm]{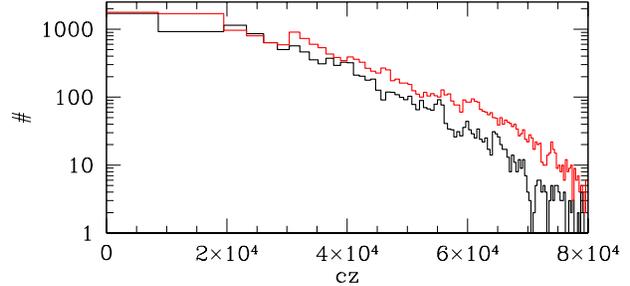}
\caption{The group number density in equal volume shells. The NGP/SGP 
subsample is represented by the lower/upper line.}
\end{figure}

\subsection{Mean group shape parameters}
We derive the projected group shape parameters using the 
moments of inertia method (cf. Carter \& Metcalfe 1983; 
Basilakos et al. 2000).
In Plionis et al. (2004) we performed a large number of Monte-Carlo
simulations to test whether random projections could result in the
observed distributions of axial ratios and we have excluded such
possibility (without this meaning that some of the groups could not be
contaminated by projections).
Of course the probability of a group being false is inversely
proportional to the group galaxy membership, $n_m$. Random
projections will affect significantly more the apparent
characteristics (dynamical and morphological) of small
groups rather than large ones 
and for this reason we have decided not to study
groups with $n_m=3$.

Based on their galaxy membership, $n_m$, we will study
separately groups with $n_m=4$, $4\le n_m< 20$ and $n_m\ge 20$,
the latter being nearer to the definition of a cluster.

The summary of the main structure parameters of the different membership
samples of groups is presented in Table 1. The
first and second columns give the group membership and the 
number of such groups respectively, the third column gives their mean redshift
while the fourth and fifth
columns show the median values of the projected axial ratio ($\bar{q}$) and
intergalaxy separation ($\bar{a}$), together with their
68\% and 32\% quantile values. 
It is evident that the considered groups are very elongated,
significantly more than what expected from random projections of field
galaxies (see Plionis et al. 2004), giving support to them being real 
dynamical entities.

\begin{table}[]
\caption[]{Summary of group subsample characteristics within 
$cz<30000$ km s$^{-1}$: $N$ is the
  number of groups, $\langle z\rangle$ is their mean redshift,
  $\bar{q}$ and $\bar{a}$ are the median values of their axial ratios and
  member intergalaxy separation.}
\tabcolsep 12pt
\begin{tabular}{cccccc} \\ \hline
 $n_m$& N  & $\langle z\rangle$   & $\bar{q}$   & $\bar{a}$ \\ \hline \hline
      \multicolumn{5}{c}{NGP} \\ \hline
 4        & 506 & 0.067 &$0.31^{+0.09}_{-0.10}$ & $0.41^{+0.09}_{-0.06}$  \\
 5-19     & 872 & 0.069 &$0.42^{+0.08}_{-0.10}$ & $0.56^{+0.12}_{-0.11}$  \\
 $\ge 20$ & 115 & 0.069 & $0.56^{+0.08}_{-0.05}$ & $1.17^{+0.27}_{-0.22}$  \\ \hline
\multicolumn{5}{c}{SGP} \\ \hline
 4         & 493 & 0.066 &$0.32^{+0.09}_{-0.10}$ & $0.37^{+0.07}_{-0.05}$  \\
 5-19      & 897 & 0.065 &$0.44^{+0.09}_{-0.09}$ & $0.49^{+0.08}_{-0.08}$  \\
 $\ge 20$  & 97  & 0.061 &$0.59^{+0.1}_{-0.06}$ & $0.92^{+0.15}_{-0.11}$  \\
\hline
\end{tabular}
\end{table}
A certain correlation is apparent
between $\bar{q}$) and $\bar{a}$ increasing with $n_m$. 
We find that the mean redshift of each group subsample is constant 
and thus the increase of the group size with $n_m$ 
cannot be due to the increase with redshift of the group linking volume
which induces the systematic trend seen in Fig.1.

The increase of the group sphericity with $n_m$
could be explained as an indication of a higher degree of
virialization, which is expected to be more rapid in systems
containing more galaxies (mass). 
Virialization processes increase the sphericity of
systems but also compactifies them from their original dispersed 
configuration, assuming that they accrete mass anisotropically along
large-scale filaments (cf. West 1994). 
The increase with $n_m$ of the group sizes, which is also accompanied with
a decrease of the galaxy density, dropping from $\sim 160$ to $\sim
20$ $h^{3}$ Mpc$^{-3}$ for the $n_m=4$
and $n_m\ge 20$ groups respectively (assuming a prolate group shape- see
next session), is quite intriguing.

Finally, comparing the 2PIGG shape parameters with
those of the UGC-SSRS2 groups (Plionis et al. 2004) we find quite
similar results. For example, the median $q$ of the $4\le n_m\le 10$
groups are $\sim 0.33 \pm 0.09$ and $\sim 0.36\pm 0.09$ for the
UGC-SSRS2 and 2PIGG samples, respectively.

\subsection{The projected axial ratio distribution} 
The derived discrete frequency distribution of the projected group axial ratio
is fitted by a continuous function using the so-called kernel
estimators (for details see Ryden 1996 and references therein)
Although we will not review this method we note that the basic 
kernel estimate of the frequency distribution is defined as:
\begin{equation}\label{eq:ker1}
\hat{f}(q)=\frac{1}{Nh} \sum_{i}^{N}\ K\left(\frac{q-q_{i}}{h}\right) \; \; ,
\end{equation}
where $q_{i}$ are the group axial ratios and
$K(t)$ is the kernel function (assumed here to be a Gaussian), 
defined so that $\int K(t) {\rm d}t=1$,
and $h$ is the ``kernel width" which determines the balance between
smoothing and noise in the estimated distribution.
The value of $h$ is chosen so that the expected value of the
integrated mean square error between the true, $f(q)$, and estimated, 
$\hat f(q)$, distributions 
is minimised (cf. Vio et al. 1994; Tremblay \& Merritt 1995).

\begin{figure}
\includegraphics[width=9.5cm]{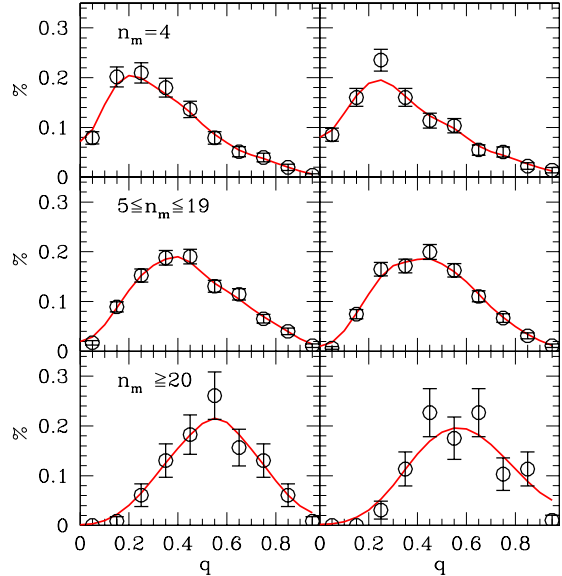}
\caption{The apparent axial ratio distributions for different group
membership. The solid line is the smooth fit from the nonparametric 
kernel estimator. {\sc Left Panel:} The NGP subsample, {\sc
Right Panel:} The SGP subsample.}
\end{figure}

In Figure 3 we present the projected axial ratio distributions for the
different membership groups (circles), as indicated in the different
panels, with their Poisson 1$\sigma$ error bars, while the 
solid lines shows the kernel estimate $\hat{f}$ for the appropriate
width, $h$. As the membership number increases, there is evidently 
a shift to less flattened systems, with the peak of the axial ratio
distribution shifting from $q\simeq 0.22$ to $\sim$0.4 and $\sim$0.55 
for groups with $n_m=4$, $5\le n_m<20$ and $n_m \ge 20$, respectively.

\subsection{True Group Shapes}
As in Plionis et al. (2004) we invert the projected axial ratio distribution
assuming that groups are either oblate or prolate spheroids.
Although there is no physical justification 
for this restriction, 
it greatly simplifies the inversion problem. Furthermore, if groups
are a mixture of
the two spheroidal populations or they have triaxial configurations
then there is no unique inversion (Plionis et al. 1991).
\begin{figure}
\includegraphics[width=9.5cm]{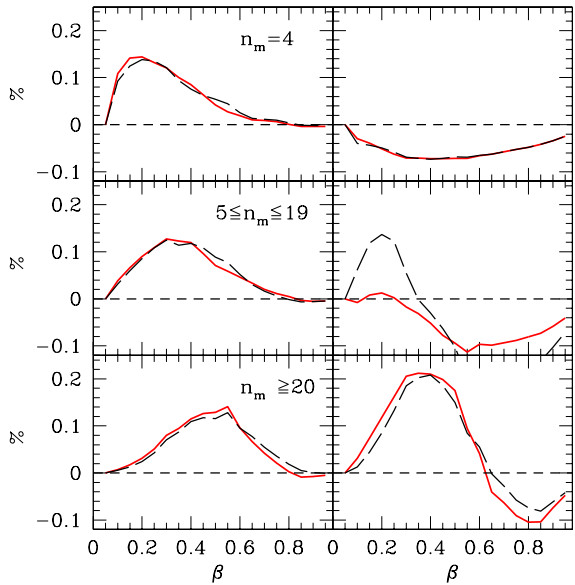}
\caption{The distribution of the intrinsic 2dFGRS group axial ratios
for the NGP (continuous line) and SGP (dashed line) subsamples
assuming that they are either prolate (left
panel) or oblate (right panel) spheroids.}
\end{figure}

Under the above restriction and the assumption that the 
orientation of groups is random
with respect to the line of sight,
the relation between the apparent and intrinsic axial ratios can be 
described by a set of integral equations, first investigated by Hubble
(1926).
Writing the intrinsic axial ratios as $\beta$ and the estimated distribution
function as $\hat N_o(\beta)$ for oblate spheroids, and $\hat
N_p(\beta)$ for prolate spheroids then the corresponding distribution
of apparent axial ratios is given for the oblate case by:
\begin{equation}\label{eq:apaobl}
\hat{f}(q)=q\int_{0}^{q}\frac{\hat{N}_{\circ}(\beta) {\rm d}\beta}
{(1-q^{2})^{1/2}(q^{2}-\beta^{2})^{1/2}}
\end{equation}
and for the prolate case by:
\begin{equation}\label{eq:apaprol}
\hat{f}(q)=\frac{1}{q^{2}}\int_{0}^{q}\frac{\beta^{2}\hat{N}_{p}(\beta) 
{\rm d}\beta}
{(1-q^{2})^{1/2}(q^{2}-\beta^{2})^{1/2}} \; \; .
\end{equation}
Inverting equations (eq.\ref{eq:apaobl}) and (eq.\ref{eq:apaprol})
gives us the distribution of real axial ratios as a function of the measured
distribution:
\begin{equation}\label{eq:oblate}
\hat{N}_{o}(\beta)=\frac{2\beta (1-\beta^{2})^{1/2}}{\pi} \int_{0}^{\beta}
\ \frac{\rm d}{{\rm d}q}\left(\frac{\hat{f}}{q} \right)\frac{{\rm d}q}
{(\beta^{2}-q^{2})^{1/2}}
\end{equation}
and
\begin{equation}\label{eq:prolate}
\hat{N}_{p}(\beta)
=\frac{2(1-\beta^{2})^{1/2}}{\pi\beta} \int_{0}^{\beta}
\ \frac{\rm d}{{\rm d}q}(q^{2}\hat{f})
\frac{{\rm d}q}{(\beta^{2}-q^{2})^{1/2}} \; \; .
\end{equation}
with $\hat{f}(0)=0$. The important point here is that
in order for $\hat{N}_{p}(\beta)$ and $\hat{N}_{o}(\beta)$
to be physically meaningful they should be positive for all
$\beta$'s. Negative values indicate that the model is unacceptable.
Integrating numerically eq.(\ref{eq:oblate})
and eq.(\ref{eq:prolate}) allowing $\hat{N}_{p}(\beta)$ and
$\hat{N}_{o}(\beta)$ to take any value, we derive the inverted
(3D) axial ratio distributions, which we present in Figure 4.

The oblate model is completely
unacceptable since it produces negative values of the inverted
intrinsic axial ratio distribution. 
Therefore, we can conclude that the 2PIGG groups shape is well
represented only
by that of prolate spheroids which is in agreement with the previous
analysis of the UGC-SSRS2 poor groups (Plionis et al. 2004) and
Shakhbazian compact groups (Oleak et al. 1995). 

The richer groups ($n_m\ge 20$) have an intrinsic axial ratio
distribution that approximates that of clusters of galaxies, as can be
seen in Fig. 5 were we compare with the APM cluster shapes
for the prolate case (which is also the best spheroidal 
model for clusters; see Plionis et al. 1991, Basilakos et al. 2000).

\begin{figure}
\includegraphics[width=9.5cm]{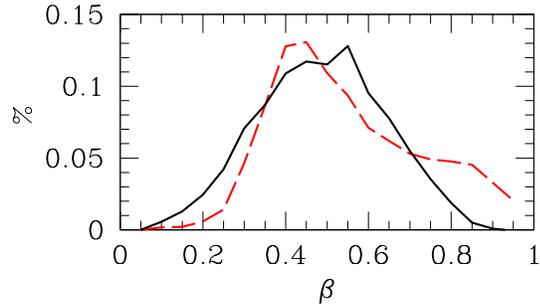}
\caption{Comparison of the intrinsic ($n_m\ge 20$) group axial ratio
  distribution (continuous line) with that of the APM cluster (dashed
  line; from Basilakos et al. 2000) for the prolate model.}
\end{figure}

\section{Conclusions}
We have measured the projected axial ratio distribution
of the 2PIGG groups of galaxies within a roughly 
volume-limited region ($cz\le 30000$
km/sec). Assuming that groups constitute a homogeneous 
spheroidal population, we numerically invert the projected 
axial ratio distribution to obtain the corresponding intrinsic 
one. The only acceptable model is that of prolate spheroids which is 
in excellent agreement with the analysis of the 
UGC-SSRS2 sample of poor groups (Plionis et al. 2004).

There is an obvious group richness-flatness relation,
seen in both projected and intrinsic axial ratio distributions,
with poorer groups being also the flatter.
For example, the peak of the projected
axial ratio distribution shifts from $q\sim$ 0.22 for groups with only
4 members to $\sim$0.55 for those with more than 19 members.
The increase of the group sphericity with richness, which extends also
to clusters,
could be explained as an indication of a higher degree of
virialization, which is expected to be more rapid in more massive
systems.

\section* {Acknowledgments}
This research was jointly funded by the European Union
and the Greek Government 
within the program 'Promotion
of Excellence in Technological Development and Research'. 
MP also acknowledges funding by the Mexican Government grant
No. CONACyT-2002-C01-39679.

\end{document}